\documentstyle[prl,preprint,aps]{revtex}
\begin{document}
\draft
\title{STOKES' DRIFT OF LINEAR DEFECTS}
\author{F. Marchesoni}
\address{Istituto Nazionale di Fisica della Materia, Universit\'a di Camerino}
\address{I-62032 Camerino (Italy)}
\author{M. Borromeo}
\address{Dipartimento di Fisica, and Istituto Nazionale di Fisica
Nucleare,}
\address{Universit\'a di Perugia, I-06123 Perugia (Italy)}
\date{Received:\today}
\maketitle
\begin{abstract}
A linear defect, viz. an elastic string, diffusing on a planar substrate traversed
by a travelling wave experiences a drag known as Stokes' drift. In the 
limit of an infinitely long string, such a mechanism is shown to be characterized by
a sharp threshold that depends on the wave parameters, the string damping constant
and the substrate temperature. Moreover, the onset of the Stokes' drift is signaled 
by an excess diffusion of the string center of mass, while the dispersion of the drifting
string around its center of mass may grow anomalous.  
\\

PCAS numbers: 05.60.-k, 66.30.Lw, 11.27.+d\\
\end{abstract}

\newpage

\section{Introduction}

Particles suspended in a viscous medium traversed by a longitudinal wave $f(kx-\Omega t)$ 
with velocity $v=\Omega /k$ are dragged along in the $x$-direction according to a
deterministic mechanism known as Stokes' drift \cite{1}. As a matter of
fact, the particles
 spend slightly more time in regions where the force acts parallel to the direction of
propagation than in regions where it acts in the opposite direction. Suppose \cite{2}
that $f(kx-\Omega t)$ represents a symmetric square wave with wavelength $\lambda=2\pi/k$
and period $T_{\Omega}=2\pi/\Omega$, capable of entraining the particles with velocity
$\pm bv$ (with $0\leq b(v) \leq 1$ and the signs $\pm$ denoting the orientation of the force).
Within one cycle, the particle velocity is positive for a longer time interval, $T_{\Omega}
/2(1-b)v$, than it is negative, $T_{\Omega}/2(1+b)v$; hence, the average drift velocity
$u\equiv b^2v$.

Recent studies \cite{2,3} have shown that thermal diffusion of the dragged particles
can affect markedly the magnitude of the Stokes' drift. A simple change of variables,
$x \rightarrow y\equiv kx-\Omega t$, makes it possible to map the Stokes' mechanism
into the more conventional problem of Brownian motion in a tilted washboard potential \cite{4};
the dependence of the drift velocity $u$ on the viscous constant $\alpha$ and the temperature 
$T$ of the suspension medium follows immediately \cite{3}. No matter what the value of 
$\alpha$, the Stokes' drift is characterized by an onset threshold that grows sharper
and sharper as $T$ is lowered toward zero; at finite temperature a residual drift occurs in 
the forbidden parameter region, due to thermal activation ({\it creep}). 
Lately, the Stokes's mechanism has been attracting the interest of an increasing number 
of experimentalists in
view of its potential applications to single-electron transport in one-dimensional 
channels \cite{5}, light storage in quantum wells \cite{6}, 
optical tweezing of colloidal particles \cite{7}, and transport by capillary waves \cite{8}, 
to mention but a few.

In this article we consider the possibility that a directed {\it linear} defect confined to 
a planar substrate can undergo Stokes' drift as an effect of a longitudinal wave travelling 
perpendicularly to its axis. By means of a convenient change of variable, 
like in Ref. \cite{3},
one expects to recast the question at hand into an instance of the more general problem
of the diffusion of an elastic string on a tilted periodic substrate \cite{9}-\cite{13}.
Such a mechanism has been used extensively in the literature to model the diffusion of 
dislocations in solids \cite{9}, the characteristic curve of Josephson junction arrays
\cite{11}, the lubrication processes taking place between slipping macroscopic 
surfaces \cite{14} and the transport of magnetic flux lines in type-II superconducting 
films \cite{15,16}.
Here, as a major difference between point-like [or (0+1)d] and linear [or (1+1)d] defects,
we show that:  (i) In the limit of infinitely long strings, creep effects and critically 
sharp Stokes' thresholds coexhist at any finite temperature; (ii) Such thresholds shift toward
lower values with increasing the temperature, due to the configurational string fluctuations 
(or phonon dressing); 
(iii) The dispersion of the drifting string around its center of mass grows anomalously with 
time.

The paper is organized as follows. In Sec. II, we determine the dependence of the drift 
thresholds on the damping constant and the temperature  of the substrate. The onset of the 
Stokes' mechanism is sharp for infinitely long strings, whereas finite-length corrections
may be computed numerically with no much effort. In Sec. III, the phenomenology of the Stokes'
 drift for linear defects
is analyzed in detail with particular attention to the excess diffusion of the string center 
of mass in the vicinity of the threshold and the anomalous dispersion of the string around its
 center of mass. In Sec. IV, we compare the effectiveness of the Stokes' drift for (0+1) and 
(1+1) dimensional defects in solids.

\section{Depinning thresholds}

As anticipated in the Introduction, the dynamics of a directed elastic string dragged by a 
longitudinal wave propagating perpendicularly to the string in its glide plane, is intimately 
related to the diffusion of the very same string on a suitable periodic substrate subjected to
 an external static bias. Therefore, we address first the problem of the depinning of a damped 
string from a tilted washboard potential \cite{10}-\cite{13} with the purpose of better 
characterizing the threshold dynamics.

The perturbed sine-Gordon (SG) equation
\begin{equation}
\label{1}
\phi_{tt} - c_0^2 \phi_{xx} + {\omega_0^2}\sin\phi =
-\alpha \phi_t + F + \zeta(x,t),
\end{equation}
represents the archetypal model of an elastic string in thermal equilibrium on a periodic
substrate \cite{17,18}. 
Here, $c_0^2$ and $\omega_0^2$ characterize the string restoring force and the depth of the 
substrate valleys, respectively; correspondingly,  $c_0$ is the maximum propagation 
(or sound) speed of a signal
along the string and $\omega_0$ is the librational (or plasma) frequency of the string itself at
the bottom of a SG valley.
The coupling of the classical field $\phi(x,t)$ to the heat bath
at temperature $T$ is reproduced here by a viscous term $-\alpha \phi_t$ and a zero-mean 
Gaussian noise source $\zeta(x,t)$. The damping constant $\alpha$ and the noise intensity
are related through the noise autocorrelation function
\begin{equation}
\label{2}
\langle\zeta(x,t) \zeta(x',t')\rangle = 2 \alpha kT \delta(t-t') \delta(x-x').
\end{equation}
As the relevant SG substrate potential $V[\phi]=\omega_0^2(1-\cos\phi)$ is tilted by the 
bias term $-F\phi$, the resulting washboard potential retains its
multistable structure only if
$|F| <F_3=\omega_0^2$ (static depinning threshold \cite{4}).

The elastic string can glide perpendicularly to the $V[\phi]$ valleys through three distinct
mechanisms:

{\it (a) Kink drift.} -- A string bridging two adjacent SG valleys bears at least one
{\it geometrical} kink $\phi_+(x,t)$, or antikink $\phi_-(x,t)$, with $\phi_{\pm}(+\infty,t)-
\phi_{\pm}(-\infty,t)=\pm 2\pi$,
\begin{equation}
\label{12}
\phi_{\pm} (x,t) =  4\arctan \left [ \exp\left( \pm 
{{x -x_0 -ut} \over {d \sqrt{1 - u^2/c_0^2}}} \right )\right ].
\end{equation}
 In the presence of a positive bias $F$, the solitons
$\phi_{\pm}$ are driven apart to $\mp \infty$, so that the string
eventually advances by one step
$2\pi$. SG solitons are known to behave like Brownian particles with finite size 
$d=c_0/\omega_0$, mass $M_0=8/d$ and average drift velocity $\pm u_F$, with  \cite{19}
$u_F=2\pi F/\alpha M_0$. This mechanism of string transport works for $F\neq 0$,
no matter how low the temperature.

{\it (b) Thermal nucleation.} -- A string segment lying along one SG valley is unstable 
against nucleation of thermal kink-antikink pairs \cite{16,20}. Thermal fluctuations 
$\zeta(x,t)$
trigger the process by activating a critical nucleus. For $F \ll F_3$ an unstable nucleus 
is well reproduced by the linear superposition of a kink and an antikink, each experiencing
two contrasting forces: a restoring force due to the vicinity of its nucleating partner and
a repulsive force due to either the tilt $F$ itself, or the preexisting (equilibrium) gas of
 kinks and antikinks \cite{18}. For $F \ll F_3$ the size of the critical nucleus, obtained by
balancing the two forces, is quite large, so that its energy is about twice the soliton rest
 energy $E_0=M_0c_0^2$. At low temperature $kT\ll E_0$, the nucleation time is thus 
exponentially long, $T_N \propto \exp(2E_0/kT)$. As a result, on establishing opposite, stationary 
kink-antikink flows, see also mechanisms (a), the string {\it creeps} in the $F$ direction with average  
center of mass velocity \cite{16} $u_{th}=2\pi n_0 u_F$, proportional to the 
density of thermal kinks (or antikinks) \cite{18}
\begin{equation}
\label{3}
n_0(T)= {1 \over d} \left ({2 \over {\pi}}\right )^{1\over 2}
\left ({{E_0} \over {kT}}\right )^{1\over 2} \exp \left(-{{E_0}\over{kT}}\right).
\end{equation}

{\it (c) Depinning from the substrate.} -- Even at low temperature, where thermal creep
is negligible, the elastic string may unlock from the SG substrate. In the overdamped regime,
$\alpha \gg \omega_0$, this happens of course for values of $F$ above the static threshold
$F_3$; in the {\it underdamped} regime $\alpha \ll \omega_0$, the depinning threshold is determined 
{\it dynamically} \cite{4}: at $T=0+$, stationary string translation is 
energetically sustainable above the new threshold
\begin{equation}
\label{4}
F_2=3.36\dots\left ( {{\alpha}\over{\omega_0}}\right)F_3.
\end{equation}
For $F>F_2$, the potential energy gained by the string during one jump
forward is larger than the energy 
dissipated, in average, by the viscous force that opposes the process.
[In the noiseless case $T\equiv 0$ \cite{11,12}, of no relevance to the
present study, the string dynamics would be characterized  by 
hysteretic  loops with upper depinning threshold at $F_3$ and lower re-pinning threshold 
$F_1=(4/\pi)(\alpha/\omega_0)F_3$.] On increasing $\alpha$ the depinning threshold
grows monotonically according to the linear law (\ref{4}), for $\alpha \ll \omega_0$, and
levels off at the constant value $F_3$ for $\alpha~ ^{ >}_{\sim} ~1.2\omega_0$.\\

For tilt values above the depinning threshold $F_2$, an underdamped string slides downhill 
with average velocity $u_{\phi}$; both the velocity of the string center of mass and the 
lateral motion of its kinks and antikinks are modulated periodically in time with period
$2\pi/u_{\phi}$. As the upper bound of $u_{\phi}$ is $F/\alpha$ \cite{4,13}, the amplitude 
of the soliton sidewise oscillations is certainly smaller than $(1/2)[2\pi /(F_2/\alpha)]c_0$, 
where $F_2$ is given by Eq. (\ref{4}) and $c_0$ is the limiting speed of any pulse propagating 
along the string, and therefore smaller than the soliton size $d$. Moreover, for $kT \ll E_0$
the mean kink-antikink distance $n_0^{-1}(T)$ is much larger than $d$, viz. $n_0(T)d \ll 1$, 
see Eq. (\ref{3}); as a consequence, the spatial configuration of the falling string is 
only negligibly modulated by its passage over the substrate barriers. The same conclusion 
applies in the opposite regime $\alpha \gg \omega_0$.

Such a periodic effect can be averaged out by adding explicitly a translational coordinate 
$X$
to the field $\phi$, namely
\begin{equation}
\label{5}
\phi(x,t) \rightarrow X(t) + \phi(x,t).
\end{equation}
On taking the spatial average
\begin{equation}
\label{6}
\langle \dots \rangle_x= \lim_{L\rightarrow \infty} {1 \over L}\int^{L/2}_{-L/2}(\dots)dx
\end{equation}
[not to be mistaken for the stochastic average $\langle \dots \rangle$ in Eq. (\ref{2})] of
both sides of Eq. (\ref{1}) along a string segment of length $L$, one obtains
\begin{equation}
\label{7}
\ddot X=-\alpha (\dot X+u_{th}) - U'(X,T) + F + {1 \over \sqrt{L}}\eta(t),
\end{equation}
where we set
\begin{equation}
\label{8}
\langle \phi_t(x,t) \rangle_x=\dot X +u_{th},
\end{equation}
\begin{equation}
\label{9}
\langle \phi_{tt}(x,t) \rangle_x=\ddot X,
\end{equation}
\begin{equation}
\label{10}
\langle \zeta(x,t) \rangle_x={1 \over \sqrt{L}} \eta(t),
\end{equation}
\begin{equation}
\label{11}
\langle V'[\phi] \rangle_x= U'(X,T).
\end{equation}
In Eqs. (\ref{8}) and (\ref{9}) we have made use of the reflection symmetry 
$\partial_t \phi_{\pm} = -\partial_t \phi_{\mp}$ of the $\phi_{\pm}(x,t)$ solutions.
Note that for $kT\ll E_0$ the $\phi_{\pm}$ center of mass $x_0$ fluctuates  \cite{19} with 
velocity variance $\langle u^2 \rangle = kT/E_0 \ll c_0^2$. In
Eq. (\ref{8}) $u_{th}$ denotes the creep velocity of the string due to the lateral drift of
kinks and antikinks, see mechanism (b) above. 

The spatial average of the random force in Eqs. (\ref{1})-(\ref{2}) is statistically
equivalent to the Gaussian noise (\ref{10}), where the overall
intensity is inverse proportional to the string length $L$, $\langle \eta(t)\rangle=0$ and
\begin{equation}
\label{13}
\langle\eta(t) \eta(t')\rangle = 2 \alpha kT \delta(t-t').
\end{equation}

The effective potential $U(X,T)$ can be computed only under certain low-temperature
approximations. Of course, it retains all the spatial symmetries of $V[\phi]$, namely
$U(-X,T)=U(X,T)$ and $U(X+2\pi,T)=U(X,T)$. In view of the rigid translation (\ref{5}), 
the identities
$\langle V'[\phi] \rangle_x \equiv  \omega_0^2 \langle\sin (X+\phi) \rangle_x = \omega_0^2 
[\sin X \langle \cos \phi \rangle_x +
\cos X \langle \sin \phi \rangle_x]$ and $\langle \sin \phi \rangle_x=0$, lead to
\begin{equation}
\label{14}
U(X,T) =  V(X)\langle \cos \phi \rangle_x
\end{equation}
with $V(X) = \omega_0^2 (1-\cos X)$. 
[Note that the averages  (\ref{6})  $\langle\cos \phi \rangle_x$ and $\langle\sin \phi \rangle_x$
are to be taken along the kink profile.]
The $T-$dependent configurational factor 
$\langle \cos \phi \rangle_x$ deviates from unity because of (i) the local $\phi$ 
steps (solitons), (ii) the extended $\phi$ excitations (phonons) sustained by thermal 
fluctuations.
The first correction is proportional to the (anti)kink density $n_0(T)$ and therefore 
exponentially small; neglecting the soliton contribution corresponds to replacing 
$\langle \cos \phi \rangle_x$ with $\langle \cos \psi \rangle_x$, where $\psi(x,t)$ represents 
the equilibrium
phonon gas of a string sitting at the bottom of one SG valley. The field $\psi(x,t)$
can be regarded \cite{17,18} as a linear superposition of phonon modes $\psi_k(x,t)$, 
with dispersion $\omega_k^2=\omega_0^2 + c_0^2 k^2$ and average amplitude set by the 
equipartition theorem $\langle \psi^2_k \rangle_x=kT/\omega_k^2$; hence
\begin{equation}
\label{15}
\langle \psi^2 \rangle_x={{kT}\over{2\pi}}\int_{-\infty}^{+\infty}{{dk}\over
{\omega_0^2 + c_0^2 k^2}}
=4{{kT}\over{E_0}}.
\end{equation}
At low temperature the distribution of the phonon field is well approximated by a
 Gaussian 
function $P[\phi]=\exp(-\psi^2/2\langle \psi^2 \rangle_x)/\sqrt{2\pi \langle \psi^2 \rangle_x}$,
so that 
\begin{equation}
\label{16}
\langle \cos \phi \rangle_x = \Re \left\{\int e^{i\psi}P[\psi]d\psi \right\}=
e^{-\langle \psi^2 \rangle_x/2}.
\end{equation}
On combining Eqs. (\ref{14}) and (\ref{16}) we conclude that
\begin{equation}
\label{17}
U(X,T) = \omega^2(T) (1-\cos X)
\end{equation}
with
\begin{equation}
\label{18}
\omega(T) = \omega_0 e^{-kT/E_0};
\end{equation}
hence, the $T$ dependence of the string depinning threshold (\ref{4}).

Let us discuss now the diffusion of the string center of mass described by the reduced 
Eq. (\ref{7}) or, after the Galileian transformation $Y=X+u_{th}t$, by
\begin{equation}
\label{19}
\ddot Y=-\alpha \dot Y - \omega^2(T)\sin(Y-u_{th}t) + F + {1 \over \sqrt{L}}\eta(t).
\end{equation}
The processes $X(t)$ and $Y(t)$  represent a driven Brownian motion on a
tilted washboard potential \cite{4}.
In the limit of an infinitely long string the effective temperature of the process
(\ref{7}),(\ref{19}) $T^*=T/\sqrt{L}$ tends to zero. This implies, for instance,
 that the depinning threshold of an underdamped string is well approximated by Eq. (\ref{4}) 
after $\omega_0$ was replaced by $\omega(T)$. As shown in Risken's textbook \cite{4}, the
depinning mechanism is abrupt; for $F=F_2(T)+$ the average drift velocity of the string 
$u_{\phi}\equiv \langle \dot X \rangle$ jumps all of a sudden from its creep value $u_{th}$
up to close to its asymptotic value $F_2(T)/\alpha$. In Eqs. (\ref{7}) and (\ref{19}) the dependence 
on the 
string temperature $T$ is restricted to the effective potential amplitude
(\ref{18}) [as a result the threshold value $F_2(T)$ gets lowered] and to the creep velocity $u_{th}(T)$. 
Note that even the dispersion
of the creep velocity tends to vanish in the limit $L \rightarrow \infty$, as the creep 
mechanism through pair nucleation is essentially a shot noise process \cite{note1}.

Finite-length corrections to the picture outlined above can be computed explicitly by applying 
the numerical recipes of Ref. \cite{4} for a finite effective temperature $T^*$. 
Depinning curves for vanishingly small $T^*$ values are displayed in Ref. \cite{13}; 
anticipation of the depinning thresholds with the string temperature is
hinted at, though not explicitly discussed, in Refs. \cite{9,10,13}. 

\section{Stokes' drift}

We go back now to the initial problem of a directed elastic string (say, a lattice dislocation)
confined to a planar substrate and subjected to a longitudinal wave $f(\phi - vt)$  
(a stress wave) propagating perpendicularly to the string direction. Let the string be 
coupled to a viscous heat bath at temperature $T$; the corresponding field equation reads  
\begin{equation}
\label{3.1}
\phi_{tt} - c_0^2 \phi_{xx}  =
-\alpha \phi_t + f(\phi -vt) + \zeta(x,t),
\end{equation}
where $\zeta(x,t)$ is defined in Eq. (\ref{2}) and $v=\Omega/k$ denotes the propagation 
speed of a travelling wave with wavelength $2\pi/k$. By means of the Galileian transformation
$\phi \rightarrow \psi \equiv \phi-vt$, Eq. (\ref{3.1}) can be rewritten as
\begin{equation}
\label{3.2}
\psi_{tt} - c_0^2 \psi_{xx}  =
-\alpha \psi_t + f(\psi) -\alpha v + \zeta(x,t).
\end{equation}
For a sinusoidal wave with $k=1$, $f(\psi) = -f^2_0\sin \psi$, this equation coincides with  
Eq. (\ref{1}) under the correspondence rule 
\begin{equation}
\label{3.3}
\omega_0 \rightarrow f_0, ~~~ F \rightarrow -\alpha v,
\end{equation}
and $\alpha, T$ and $c_0$ unchanged.

\subsection{Onset thresholds}

The center of mass of the string $\psi(x,t)$ diffuses according to
Eqs. (\ref{7}) and (\ref{19}), after 
transformation (\ref{3.3}). As long as $\psi(x,t)$ is locked to the bottom of a (creeping)
SG valley, see Eq. (\ref{19}), the initial string $\phi(x,t)$ drifts with velocity $u_{\phi}\simeq v$; upon 
depinning, the center of mass of $\psi(x,t)$ travels to the left with negative velocity 
$u_{\phi} \simeq F/\alpha=-v$; correspondingly, $\phi(x,t)$ gets traversed by the propagating 
wave without undergoing any appreciable drag, were not for the residual creep velocity $u_{th}$.

The relevant Stokes' threshold $v_S$ admits two simple limiting approximations; in the 
overdamped regime $\alpha \gg \omega_0$
\begin{equation}
\label{3.4}
v_S(\alpha \rightarrow \infty)=v_3 = {{F_3}\over {\alpha}}={{f^2(T)}\over {\alpha}},
\end{equation}
in the underdamped regime $\alpha \ll \omega_0$
\begin{equation}
\label{3.5}
v_S(\alpha \rightarrow 0)=v_2 = {{F_2}\over {\alpha}}=3.36\dots f(T),
\end{equation}
where $f(T)=f_0\exp(-kT/E_0)$. Stokes' drift requires that $v<v_S$, i.e. travelling waves
 must be selected with sufficiently low frequency or short wavelength. At low damping the 
threshold velocity $v_S \simeq v_2$ is insensitive to $\alpha$ and the drift effect is more 
pronounced. In the limit of infinitely long strings, $L \rightarrow \infty$, the threshold 
mechanism grows sharper and sharper, as to be expected for the diffusion of a Brownian 
particle 
in a tilted washboard potential at zero temperature $T^*$. At variance with the (0+1) 
dimensional case \cite{2,4}, the onset of the Stokes' drift for an elastic string in thermal 
equilibrium is thus a {\it critical transition}, even at finite (but relatively low) temperature.

\subsection{Excess diffusion}

As proposed in Ref. \cite{21}, a dynamic threshold can be characterized in terms of the 
diffusion coefficient
 \begin{equation}
\label{3.6}
D=\lim_{t\rightarrow \infty} {{\langle X^2(t) \rangle-\langle X(t) \rangle^2}\over{2t}}
\end{equation}
of the process at hand. Here, the function $D(v)$ tends to an exponentially small value 
$D(0) \propto \exp(-kT^*/f(T)^2)$
 for $v \rightarrow 0$, and to the Einstein constant $D(\infty)=kT^*/\alpha$ for
 $v \rightarrow \infty$ \cite{4}. Both limits vanish for an infinitely long string, 
$T^*=T/\sqrt{L} \rightarrow 0$. More notably, in the vicinity of the Stokes' threshold $v_S$,
$D(v\simeq v_S)$ is expected to shoot up drammatically; such an effect, know as {\it excess}
diffusion, may be enhanced either by lowering the temperature or by weakening the viscous coupling,
independently. In the overdamped regime (better said, in the Smoluchowski limit 
$\alpha \rightarrow \infty$) and for $T^* \rightarrow 0$, the diffusion constant $D(v_3)$ 
decays much slower than $D(\infty)$, that is  \cite{22}
\begin{equation}
\label{3.66}
{{D(v_3)}\over {D(\infty)}} \simeq  G_0\left [ {{f^2(T)}\over {kT^*}}
\right ]^{2\over 3} \propto L^{{1\over 3}}, 
\end{equation}
with $G_0\simeq 0.90$. In the underdamped
 regime, the ratio $D(v_2)/D(\infty)$ turns out to grow even faster,
 although an 
appropriate scaling law has not 
been determined, yet \cite{21}. In both instances
the excess diffusion peak is predicted to be centered very narrowly around the relevant 
threshold velocity $v_S$.

\subsection{Anomalous dispersion}

Away from the Stokes' threshold $v_S$, the string either drifts along with the travelling 
wave, if $v<v_S$, or creeps forward in the same direction, but with marginal velocity
$u_{th}\propto v\exp(-E_0/kT)$, if $v>v_S$. The spatial configuration of $\phi(x,t)$, or
equivalently of $\psi(x,t)$, is markedly different in the two opposite dynamical regimes.
The {\it creeping} string $\phi(x,t)$ corresponds to the case of the string $\psi(x,t)$
falling down a tilted periodic substrate (\ref{3.2}) ($v>v_S$): As shown in Ref. \cite{13},
under stationary periodic boundary conditions the dispersion
\begin{equation}
\label{3.7}
S(t)= \langle \langle \psi^2(x,t)\rangle_x-\langle \psi(x,t)\rangle_x^2\rangle, 
\end{equation}
of an unlocked string segment of finite length $L$ is insensitive to the substrate, 
$S(\infty)= kT L/4 c_0^2$.  
Accordingly, the spatial dispersion of a creeping string with $L
\rightarrow \infty$ is expected to be the same as for a directed elastic
string diffusing in a plane, {\it free} of periodic modulations, either
in space or in time, i.e.
\begin{equation}
\label{3.8}
S(t)= {{kT}\over {c_0}}\left (  {t\over{\pi \alpha }}\right )^{{1 \over 2}}. 
\end{equation}

The counterpart of a {\it drifting} string $\phi(x,t)$ is a string $\psi(x,t)$ locked to the
tilted substrate (\ref{3.2}); therefore, the spatial configurations of both strings result in a 
linear superposition of diffusing kinks and antikinks of size $d=c_0/f_0$, separated by an 
average distance $n_0^{-1}(T)$. Here the soliton width depends on the wave amplitude 
$f_0$ and on
the string elastic constant $c_0$, while the soliton height concides with the wavelength 
$2\pi$ (for $k=1$). The question as how the dispersion $S(t)$ of an infinitely long string pinned to 
its SG substrate evolves with time, due to thermal pair nucleation, has been addressed in 
Ref. \cite{23}: Under the further assumption that kink-antikink collisions are destructive, 
namely for \cite{16}
\begin{equation}
\label{3.9}
v \leq 2 \left ({{2\alpha}\over{f_0}} \right )^2v_3,
\end{equation}
the {\it creep dispersion grows anomalous} with fractional exponent
larger than 1/2, that is
\begin{equation}
\label{3.10}
S(t) \simeq 0.32 (u_{th}t)^{2/3}.
\end{equation}
Condition (\ref{3.9}), clearly satisfied in the overdamped regime, may apply to lower 
damping values, too, provided that $v$ is taken small enough. 

In conclusion, while for $v>v_S$ the string $\phi(x,t)$ is almost {\it transparent} to the 
travelling wave $f(\phi-vt)$, for $v<v_S$
the string spreads out anomalously with exponent 2/3 and its center of mass drifts with 
velocity
$u_{\phi}$ close to $v$.

\section{Conclusions}

Longitudinal waves in solids may drag along point-like and linear
defects, alike, a mechanism known as Stokes' drift. The onset of such a
transport process, though, exhibits distinct properties depending on the
dimensionality of the movable defects involved:\\
(i) The depinning threshold is very sharp in the case of a {\it long}
defect line, no matter what the substrate temperature. Increasing the
temperature may push indeed the Stokes' threshold $v_S$ toward lower values,
but causes no broadening of the activation region, at variance with the
depinning of
point-like defects. In view of this remark, we conclude that Stokes' drift of linear
defects is characterized by a clear-cut threshold signature.\\
(ii) Due to their spatial extension, linear defects subjected to Stokes'
drift undergo a substantial deformation in the form of an enhanced
transversal dispersion. Moreover, a dragged defect line acquires a
multi-kink profile, where the geometry of each individual (anti)kink
component depends on the incoming wave parameters. The ensuing internal
dynamics of the transported linear defects adds to the dissipative properties of the
propagation medium overall \cite{last}.


\begin{references}
\bibitem{note1} In Eq. (\ref{7}), fluctuations of $u_{th}$ would contribute to the intensity of
the noise source $\eta(t)/\sqrt{L}$ a term proportional to $n_0(T)d$, i.e. negligible at low temperatures \cite{19}.
\end{references}
\end{document}